\documentclass[twocolumn]{aastex62}

\usepackage[utf8]{inputenc}

\graphicspath{{./}{figures/}}

\received{---}
\revised{---}
\accepted{\today}
\submitjournal{ApJ}

\shorttitle{Young radio sources in the cores of GRG}
\shortauthors{Bruni et al.}
\begin{document}

\title{A DISCOVERY OF YOUNG RADIO SOURCES IN THE CORES OF \\ 
GIANT RADIO GALAXIES SELECTED AT HARD X-RAYS}

\correspondingauthor{G. Bruni}
\email{gabriele.bruni@inaf.it}

%%%%%%%%%%%%%%%%%%%%%%%%%%% AUTHORS %%%%%%%%%%%%%%%%%%%%%%%%%%%%%%%%%

\author[0000-0002-5182-6289]{G. Bruni}
\affil{INAF - Istituto di Astrofisica e Planetologia Spaziali, via Fosso del Cavaliere 100, 00133 Roma, Italy}
 \author[0000-0003-0543-3617]{F. Panessa}
 \affil{INAF - Istituto di Astrofisica e Planetologia Spaziali, via Fosso del Cavaliere 100, 00133 Roma, Italy}
 \author{L. Bassani}
 \affil{INAF - Osservatorio di Astrofisica e Scienza dello Spazio di Bologna, Via Piero Gobetti 93/3, 40129 Bologna, Italy}
 \author[0000-0002-4090-1327]{E. Chiaraluce}
 \affil{INAF - Istituto di Astrofisica e Planetologia Spaziali, via Fosso del Cavaliere 100, 00133 Roma, Italy}
 \affil{Dipartimento di Fisica, Università di Roma Tor Vergata, via della Ricerca Scientifica 1, 00133 Roma, Italy}
 \author[0000-0002-4184-9372]{A. Kraus}
 \affil{MPIfR - Max Planck Institute for Radio Astronomy, auf dem H\"ugel 69, 53121 Bonn, Germany}
 \author[0000-0003-1246-6492]{D. Dallacasa}
 \affil{DiFA - Dipartimento di Fisica e Astronomia, Università di Bologna, via P. Gobetti 93/2, 40129 Bologna, Italy}
 \affil{INAF - Istituto di Radioastronomia, via Piero Gobetti 101, 40129 Bologna, Italy}
 \author[0000-0002-2017-4396]{A. Bazzano}
 \affil{INAF - Istituto di Astrofisica e Planetologia Spaziali, via Fosso del Cavaliere 100, 00133 Roma, Italy}
  \author[0000-0002-8606-6961]{L. Hern\'andez-Garc\'ia}
 \affil{IFA - Instituto de F\'isica y Astronom\'ia, Facultad de Ciencias, Universidad de Valpara\'iso, Gran Breta{\~n}a 1111, Playa Ancha, Valpara\'iso, Chile}
 \author{A. Malizia}
 \affil{INAF - Osservatorio di Astrofisica e Scienza dello Spazio di Bologna, Via Piero Gobetti 93/3, 40129 Bologna, Italy}
\author{P. Ubertini}
\affil{INAF - Istituto di Astrofisica e Planetologia Spaziali, via Fosso del Cavaliere 100, 00133 Roma, Italy}
\author{F. Ursini}
\affil{INAF - Osservatorio di Astrofisica e Scienza dello Spazio di Bologna, Via Piero Gobetti 93/3, 40129 Bologna, Italy}
\author{T. Venturi}
\affil{INAF - Istituto di Radioastronomia, via Piero Gobetti 101, 40129 Bologna, Italy}



%%%%%%%%%%%%%%%%%%%%%%%%%%%% ABSTRACT %%%%%%%%%%%%%%%%%%%%%%%%%%%%%%%%%%%%%%%%%%

\begin{abstract}

Giant Radio Galaxies (GRG) are the largest single entities in the Universe, having a projected linear size exceeding 0.7 Mpc, which implies that they are also quite old objects. They are not common, representing a fraction of only  $\sim$6\% in samples of bright radio galaxies. While a census of about 300 of these objects has been built in the past years, still no light has been shed on the conditions necessary to allow such an exceptional growth, whether of environmental nature or linked to the inner accretion properties. Recent studies found that samples of radio galaxies selected from hard X-ray AGN catalogs selected from {\emph{INTEGRAL}}/IBIS and {\emph{Swift}}/BAT (thus at energies $>$20 keV) present a fraction of GRG four times larger than what found in radio-selected samples.

We present radio observations of 15 nuclei of hard X-ray selected GRG, finding for the first time a large fraction (61\%) of young radio sources at the center of Mpc-scale structures. Being at the center of GRG, these young nuclei may be undergoing a restarting activity episode, suggesting a link between the detected hard X-ray emission - due to the ongoing accretion - and the reactivation of the jets. %Together with the large abundance of GRG found among hard X-ray selected radio galaxies, this may suggest a scenario in which multiple nuclear radio phases, relatively close in time, favor the growth of radio galaxies up to the Mpc size typical of GRG. 

\end{abstract}



%%%%%%%%%%%%%%%%%%%%%%%%%%%%%%%% TEXT %%%%%%%%%%%%%%%%%%%%%%%%%%%%%%%%%%%%

\section{Introduction} \label{sec:intro}

A fraction of radio galaxies ($\sim$6\% in the 3CR catalogue, \citealt{1999MNRAS.309..100I}) exhibits exceptional projected linear extents, i.e. above 0.7 Mpc, making them the largest individual objects in the Universe. Both Fanaroff-Riley type I and type II radio galaxies (FRI and FRII respectively, \citealt{1974MNRAS.167P..31F}) are represented in samples of Giant Radio Galaxies (GRG). While FRI GRG are associated with early type galaxies, those with FRII morphology are hosted both in early type galaxies and quasars. Despite the overall uncertainties underlying the assumption that spectral ages of radio galaxies are representative of their dynamical ages (\citealt{1999A&A...344....7P}), it seems plausible that GRG are, on average, very old radio sources, with radiative ages between 10$^7$-10$^8$ years \citep{2004AcA....54..249M, 2008MNRAS.385.1286J}. Nevertheless, it is still under debate whether their extreme size is due to the core/accretion properties, possibly with more than one activity episode (\citealt{1996MNRAS.279..257S}), to a convenient environment conditions, with GRG expanding in a low density InterGalactic Medium (IGM, \citealt{2004evn..conf..117M}), or to a combination of both effects. A more recent study using a population of hundreds of objects, and spanning different cosmic epochs (0.016$<z<$3.22), seems to exclude a main role for IGM in the expansion of the lobes (\citealt{2018ApJS..238....9K}), failing in finding a significant match of GRG with cosmic voids.   

On the other end of radio galaxies evolution, nearly 8.5$\div$10\% of radio sources from surveys at kpc-scale resolution present a spectral shape peaking in the MHz-GHz frequency range (\citealt{1998PASP..110..493O}). Considered to be the young precursors of FRI and FRII radio galaxies, they were divided into three classes depending on the synchrotron peaking range: Compact Steep Spectrum sources (CSS, \citealt{1990A&A...231..333F}) peaking at frequencies $<$500 MHz, GigaHertz Peaked Spectrum sources (GPS, \citealt{1991ApJ...380...66O}), peaking between 500 MHz and a few GHz, and High Frequency Peakers (HFP, \citealt{2000A&A...363..887D}), peaking at frequencies higher than $\sim$5 GHz. An anticorrelation between linear size and peaking frequency was found for the three groups, due to synchrotron self absorption, and spanning a scale between 1 and 20 kpc for CSS, less than 1 kpc for GPS, and less than 100 pc for HFP (\citealt{1998PASP..110..493O,2003PASA...20...79D}). This suggests an evolutionary track from HFP to CSS, due to the expansion of plasma within the InterStellar Medium (ISM) of the host galaxy. Further support to this scenario was given by Very Long Baseline Interferometry (VLBI) observations with a pc-scale angular resolution, revealing typical morphologies of twin lobed objects, sometimes with a weak central nucleus, similarly to what is found for larger and older radio galaxies. Jets and nuclei are more prominent in compact steep spectrum quasars (\citealt{1980ApJ...236...89P,1994ApJ...432L..87W, 1997A&A...325..943S, 2006A&A...450..959O}). 
Estimates of the kinematic age for the most compact sources have revealed an age of below $\sim$10 kyr (\citealt{1998A&A...337...69O}), as also confirmed by radiative aging estimates (\citealt{1999A&A...345..769M}). Another long-debated  scenario, that eventually did not find sufficient support from observations, tried to explain the compact size as a consequence of a dense medium, frustrating the expansion of the jets (\citealt{1984AJ.....89....5V,1997ApJ...485..112B}). 

Finding one of these sources at the center of a GRG, would imply that the core is undergoing an event that re-triggered the radio activity, since the Mpc-scale structure suggests a previous expansion phase lasted a few hundreds of Myr at least. With this work, we aim at testing the presence of young radio components in the core of a hard X-ray selected sample of GRG, in order to estimate the fraction of such objects with signature of restarting radio activity.

%%%%%%%%%%%%%%%%%%%%%%%%%%%%%%%%%%%%%%%%%%%%%%%%%%%%%%%%%%%%%%%%%%%%%%%%%%%%%%%%

\section{The hard X-ray selected GRG sample}

This work is part of our ongoing multi-wavelength study of a sample of hard X-ray selected GRG. \cite{2016MNRAS.461.3165B} performed a combined radio/hard X-ray study of the AGN population selected in the hard X-ray band by  {\emph{INTEGRAL}}/IBIS and {\emph{Swift}}/BAT surveys, finding 64 objects matching radio galaxies with known redshift. NVSS and SUMSS maps allowed an estimate of the projected linear size, and it turned out that 14 of them (i.e. $\sim$25\% of the sample) can be classified as GRG. Considering the overall fraction of giant sources in samples of radio galaxies - e.g. $\sim$6\% in the 3CR catalogue (\citealt{1999MNRAS.309..100I}), 8\% in the 3CRR sample (\citealt{1983MNRAS.204..151L}) - this fraction is impressively large, and suggests a connection between the nuclear properties and the evolution of the radio structure up to the Mpc scale. Nevertheless, the same authors do not exclude that an observational bias, due to the hard X-ray selection, could affect the GRG fraction estimate in their work.  

Since 2014, we have been carrying out an observing campaign on these 14 GRG, aiming at testing the physical conditions in these peculiar sources. From GMRT data at 325 MHz on a subsample of 4 objects, we could find two newly discovered GRG (IGR J14488-4008, \citealt{2015MNRAS.451.2370M}; IGR J17488-2338, \citealt{2014A&A...565A...2M}). From a VLBI follow-up, another source showed clear signs of an extreme re-orientation (PBC J2333.9-2343, \citealt{2017A&A...603A.131H}), also possibly connected to a restarted activity. Finally, \cite{2018MNRAS.481.4250U} presented a study of the X-ray versus radio luminosity properties: interestingly enough, the radio luminosity of the lobes is lower than what expected from the core one, assuming that the actual core luminosity was the same for all the source lifetime. This may suggest a recently restarted nuclear activity. These results pointed towards the importance of a study of the core components of these GRG, aiming at finding possible differences with respect to common radio galaxies.
In this work, we present the radio spectra from 150 MHz to 10 GHz of the 14 GRG of the sample, plus an additional source (B2 1144+35B) from the latest {\emph{INTEGRAL}}/IBIS AGN catalog (\citealt{2016MNRAS.460...19M}). 

%%%%%%%%%%%%%%%%%%%%%%%%%%%%%%%%%%%%%%%%%%%%%%%%%%%%%%%%%%%%%%%%%%%%%%%%%%%%%%%%

\section{Radio observations and ancillary data} \label{sec:data}

In the following, we present radio flux densities for the cores of our GRG sample collected during our observing campaign at the Effelsberg-100m single dish, as well as from previous surveys, archives, and works in the literature.

\subsection{Effelsberg-100m single dish observations}
  
Observations with the Effelsberg-100m single dish were conducted from March to July 2018
with three secondary focus receivers: S60mm, S36mm and S28mm - see Tab. \ref{Effelsberg} 
for details.
At the lowest frequency (4.8 GHz) the observations were done with On-The-Fly (OTF) maps, in order to 
be able to discriminate the core emission from the extended structure during data analysis.
At 8.5 and 10.5 GHz the sources were observed by Cross-Scans over the core position (in 
azimuth and elevation), conveniently choosing the position angle in order to avoid the jets 
and lobes emission. 
Data reduction was performed with the {\tt TOOLBOX2}\footnote{\href{https://eff100mwiki.mpifr-bonn.mpg.de/doku.php}{https://eff100mwiki.mpifr-bonn.mpg.de/doku.php}} software; for the reduction of the OTF
maps, the {\tt NOD3} package was used in addition (\citealt{2017A&A...606A..41M}). 
In case of the maps, the first step of the data reduction was a baseline correction and the
application of the {\tt Restore} algorithm to combine the data of the two receiver pixels (feeds). The antenna temperature of the cross-scans were determined by a
Gaussian fit to the observed pattern, a correction for the (usually small) pointing offset and by averaging the amplitudes from all subscans. 
In both cases, further corrections were made for atmospheric attenuation and the gain-elevation effect of the antenna. Finally, the observed antenna temperatures were converted into the absolute flux density scale by comparing with observations of suitable calibrators like 3C286 etc. (using the scale of \citealt{1977A&A....61...99B}). Finally, we extracted the core flux density by mean of a single-component Gaussian fitting on the images.

%%%%%%%%%%%%%%%%%%%%%%%%
\begin{deluxetable}{llll}
\tablecaption{Used receivers of the Effelsberg-100m telescope. \label{Effelsberg}}
%\tablewidth{700pt}
%\tabletypesize{\scriptsize}
\tablehead{
\colhead{} 	& 
\colhead{S60mm} 	& 
\colhead{S36mm} 	& 
\colhead{S28mm} 	}
%%%%%%%%%%%%%%%%%%%%%%%%
\startdata
Center frequency			& 4.85 GHz	& 8.35 GHz	& 10.45 GHz	\\
Typical $T_{sys}$ (zenith)		& 25 K		& 22 K		& 50 K		\\
HPBW				& 145 arcsec	& 82 arcsec	& 66 arcsec	\\
Sensitivity				& 1.55 K/Jy	& 1.35 K/Jy	& 1.35 K/Jy	\\
N. of pixels			& 2			& 1			& 2			\\
\enddata
\end{deluxetable}

\subsection{Data from surveys and archives}

In addition to the Effelsberg data, we also collected measurements from surveys at lower frequencies. In particular, we made use of the recently published TIFR GMRT Sky Survey at 150 MHz (TGSS, \citealt{2017A&A...598A..78I}), the Westerbork Nothern Sky Survey at 325 MHz (WENSS, \citealt{1997A&AS..124..259R}), the Sydney University Molonglo Sky Survey at 843 MHz (SUMSS, \citealt{2003MNRAS.342.1117M}), and the NRAO VLA Sky Survey at 1.4 GHz (NVSS, \citealt{1998AJ....115.1693C}). Due to the extended emission at low frequencies, the different angular resolution can be an issue when trying to measure the core flux density: in general, we refer to the ``core region" as the unresolved central component visible at the centre of each radio galaxy on scales of tens of arcsec at most, and that can be easily disentangled from the remainder of the radio source (i.e. lobes and hot-spots). This becomes more and more difficult at low frequencies given the progressively worse angular resolution and the generally flat or even inverted spectrum of such components. Thus, we only considered the data when radio sources were resolved enough to allow a good core identification, also comparing WENSS, TGSS and NVSS. At higher frequencies the core dominates the emission, thus this is not an issue. For one southern source (PKS 2014$-$55) we report the SUMSS flux density for the core region, although we do not consider it when building the spectral shape, since the core is known to be resolved at higher resolution with ATCA (see subsection \ref{literature}). 

We also browsed the NRAO VLA Archive Survey (NVAS, \citealt{2008SPIE.7016E..0OC}) looking for archive images of our sources, and finding data for two of them: 4C +63.22 (23.4$\pm$1.1 mJy at 1.4 GHz, and 14.3$\pm$0.7 mJy at 4.5 GHz) and 4C +34.47 (338$\pm$17 mJy at 4.8 GHz, and two epochs at 15 GHz with 767$\pm$38 mJy and 226$\pm$11 mJy). Finally, for Mrk 1498, we analyzed also the JVLA archive data from 2015: since the full bandwidth of these observations was 4 GHz (spanning the observing frequency from 4 GHz to 8 GHz), we split that into four measurements of 1 GHz bandwidth each, in order to increase the frequency coverage. The resulting flux densities are: 130$\pm$13 mJy at 4.5 GHz, 127$\pm$13 mJy at 5.5 GHz, 125$\pm$12 mJy at 6.5 GHz, 124$\pm$12 mJy at 7.5 GHz. A dedicated paper on this source, presenting these JVLA data in more details, is in progress (Hern\'andez-Garc\'ia et al., in prep.).

\subsection{Data from the literature}
\label{literature}
\begin{figure*}[tbp]
\begin{center}
\includegraphics[scale=0.60,angle=0]{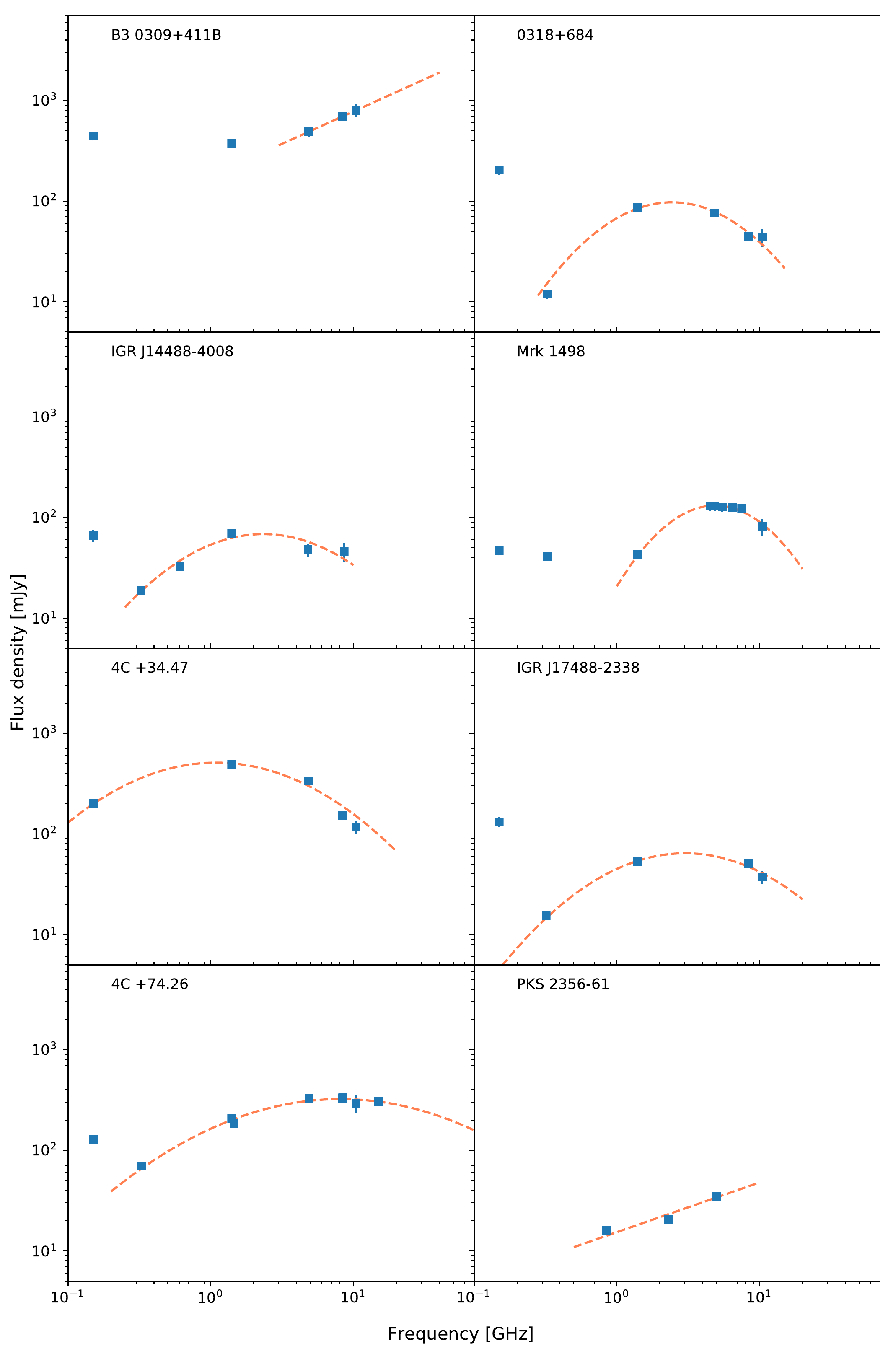}
\caption{GPS radio cores spectra. Errors in logarithmic scale are smaller than symbol size for most of measurements. The point at 250 GHz for 4C +74.26 is out of scale. \label{fig:1}}
\end{center}
\end{figure*}

For six sources we could also find data in the literature, at a suitable angular resolution to discriminate the core from the extended structure. We used the data to improve the spectral coverage and ease flux density peak identification. These are:

\begin{itemize}

\item \emph{PKS 0707-35}: the source was previously studied by \cite{2013MNRAS.436..690S}, who presented ATCA images at 1.4 GHz and 2.3 GHz. The flux densities reported for the core are 44.6 mJy and 53 mJy, respectively, with an inverted spectral index $\alpha$=0.44 (adopting the convention $S=\nu^\alpha$, where $S$ is the flux density, $\nu$ the frequency, and $\alpha$ the spectral index). We use these two points, together with TGSS observations, to draw the spectral shape from 150 MHz to 2.3 GHz, finding no peak in this range. Given the low declination of this source, it was not possible to observe it from Effelsberg, and no data from the main radio surveys are present.  

\item \emph{B2 1144+35B}: multi-epoch VLA data, from 1.4 GHz to 43 GHz, for the arcsec-core of this target have been presented in \cite{1999ApJ...522..101G, 2007A&A...474..409G}. A substantial flux density variability has been found, with a steady brightening from 1974 to 1992, followed by a dimming at all frequencies from 1992 to 2006. The authors suggest that the interaction with the external medium could be the most probable scenario to explain the variability over 3 decades. In figure \ref{fig:1} we present data for all the available VLA epochs with more than two frequencies, plus the ones taken in Effelsberg showing again flux densities lower than the previous ones, confirming the decreasing trend. In addition to that, we report also TGSS and NVSS data, the latter being taken when the core flux density was at its relative maximum.

\item \emph{IGR J14488-4008}: we observed this source at 325 MHz and 610 MHz during our GMRT campaign in 2014, which results were partially published (only the 610 MHz image) in \cite{2014A&A...565A...2M} Here we report the two flux densities (18.8$\pm$1.9 mJy at 325 MHz; 32.3$\pm$1.3 mJy at 610 MHz), the first one  being extracted from the corresponding map to be presented in a future paper (Bruni et al. in prep.). In addition to this, we also report the flux densities at 5 and 8 GHz from the ATPMN southern emisphere survey, carried out with ATCA (\citealt{2012MNRAS.422.1527M}).

\item \emph{PKS 2014-55}: Flux densities at 4.8 GHz and 8.6 GHz from ATCA observations (25.21$\pm$0.07 mJy and 20.2$\pm$0.05 mJy, respectively) for this source were provided by L. Saripalli (private communication), following the work presented in \cite{2007mru..confE.130S}. These values, together with the SUMSS measurements presented in table \ref{fluxes}, are the only data available for this source. 

\item \emph{4C +74.26}: \cite{1992MNRAS.259P..13P} presented a multi-frequency VLA study of this source's nucleus, from 0.3 to 15 GHz, plus an additional point at 270 GHz from \cite{1989MNRAS.236P..13R}. A clear GPS spectrum is visible, peaking at about 8 GHz. Previous measurements from the literature are fully compatible with the ones collected during our campaign. In addition to these, we also plot the point at 150 MHz from TGSS, that shows a turn-up of the spectrum at low frequencies.  

\item \emph{PKS 2331-240}: a study of this source was presented in \cite{2017A&A...603A.131H, 2018MNRAS.478.4634H}, showing that a Blazar-like nucleus is present, implying a dramatic change of the jet axis with respect to the GRG structure likely on the plane of the sky. In Fig. \ref{fig:1} we report the data presented in that work, in addition to the ones collected here. The flux density discrepancy at low frequencies ($\sim$150 MHz) is due to the better angular resolution of TGSS with respect to GLEAM. As suggested by those authors, restarting activity is the most probable scenario for this source, although there is no evidence of the time when it happened. This is in agreement with \cite{2017ApJ...836..174C}, that presented the source as restarting, given the convex radio spectrum (see Sec. \ref{sec:GPS} for a discussion of their work).

\item \emph{PKS 2356-61}: this source has a too low declination for Effelsberg observations. We present the measurement at 1.4 GHz, published in \cite{2018MNRAS.481.4250U}, and two ATCA measurements at 2.3 GHz and 5 GHz from \cite{1993MNRAS.263.1023M, 1997MNRAS.284..541M}, of 20.5$\pm$2.0 mJy and 35.0$\pm$3.5 mJy, respectively.

\end{itemize}

%%%%%%%%%%%%%%%%%%%%%%%%
\begin{deluxetable*}{lccccccc}
\tablecaption{List of collected flux densities (in mJy) at different frequencies (in GHz) for the 15 GRG of our sample. First three columns are measurements from TGSS, WENSS, and NVSS surveys, respectively, except starred values at 325 MHz (flux densities from our GMRT campaign), and daggered values at 1.4 GHz (measurement from the literature for PKS 0707-35 and PKS 2356-61, while from SUMSS at 843 MHz for PKS 2014-55). The second three columns are from our Effelsberg-100m campaign. The last column reports the estimated peak frequency in GHz for sources with a GPS spectral shape. See section \ref{literature} for additional measurements from the literature. \label{fluxes}}
%\tablewidth{700pt}
%\tabletypesize{\scriptsize}
\tablehead{
\colhead{} &
\colhead{} &
\colhead{} &
\colhead{} &
\multicolumn{3}{c}{Effelsberg-100m} &
\colhead{} \\
\cline{5-7}
\colhead{Source ID} 		&
\colhead{$S_{0.150}$} 		&
\colhead{$S_{0.325}$} 		&
\colhead{$S_{1.4}$} 		&
\colhead{$S_{4.8}$} 		& 
\colhead{$S_{8.5}$} 		& 
\colhead{$S_{10.5}$} 		&
\colhead{$\nu_{\rm peak}$}\\
\colhead{} 			& 
\colhead{(mJy)} 	& 
\colhead{(mJy)} 	& 
\colhead{(mJy)} 	& 
\colhead{(mJy)} 	& 
\colhead{(mJy)} 	& 
\colhead{(mJy)} 	&
\colhead{(GHz)}			
%
%\colhead{} &
%\colhead{(GHz)} &
%\colhead{(GHz)} &
%\colhead{(GHz)} &
%\colhead{(GHz)} &
%\colhead{(GHz)} &
%\colhead{(GHz)} &
%\colhead{(GHz)} 
%%%%%%%%%%%%%%%%%%%%%%%%
} 
\startdata
%--------------------------------------------------------------------------------------------------------------------------------------------------------------------------------------------------------------------------------
%				0.150			0.325				1.4					4.8				8.5				10.5				Peak
%--------------------------------------------------------------------------------------------------------------------------------------------------------------------------------------------------------------------------------
B3 0309+411B		& 446$\pm$45 		& \nodata 				& 374$\pm$37 			& 489$\pm$49 		& 694$\pm$15 		& 801$\pm$111 	& $>$10	\\
J0318+684     		& 204$\pm$20 		& 12$\pm$1 		& 86.8$\pm$8.7 		& 76$\pm$8  		& 45$\pm$2 		& 44$\pm$9 		& 2.4		\\
PKS 0707-35         	& 45$\pm$5 		& \nodata 				& 44$\pm$4$^\dagger$	& \nodata 			& \nodata 			& \nodata 			& --		\\   
4C 73.08			& \nodata			& \nodata				& 14.5$\pm$1.6 		& 10$\pm$1 		& \nodata			& 11$\pm$2 		& --		\\
B2 1144+35B		& 368$\pm$37 		& 658$\pm$66			& \nodata				& \nodata			& 169$\pm$4		& 164$\pm$24		& --		\\
HE 1434-1600   	& \nodata			& \nodata				& 75.0$\pm$7.5		& 56$\pm$3		& 61$\pm$1		& 55$\pm$8		& --		\\
IGR J14488-4008     & 66$\pm$9 	& 18.8$\pm$1.9$^\star$	& 69.7$\pm$7.0		& \nodata			& \nodata			& \nodata			& 2.3		\\        
4C +63.22           	& \nodata			& \nodata				& 10.6$\pm$0.1		& \nodata			& \nodata			& \nodata			& --		\\   
Mrk 1498  		& 47$\pm$5		& 41$\pm$4			& 43$\pm$4			& 130$\pm$13  	& \nodata			& 81$\pm$16		& 4.9		\\
4C +34.47           	& 203$\pm$20		& \nodata				& 493$\pm$49			& 338$\pm$17		& 153$\pm$3		& 117$\pm$17		& 1.0		\\   
IGR J17488-2338	& 132$\pm$15		& 15.5$\pm$1.6$^\star$	& 53.2$\pm$5.3		& \nodata			& 51$\pm$1		& 37$\pm$5		& 3.0		\\
PKS 2014-55 		& \nodata			& \nodata				& 389$\pm$12$^\dagger$	& \nodata			& \nodata			& \nodata			& --		\\
4C +74.26           	& 129$\pm$13		& \nodata				& 209$\pm$21 			& \nodata			& 328$\pm$13		& 294$\pm$59		& 8.1		\\   
PKS 2331-240        	& 366$\pm$37		& \nodata				& 801$\pm$80 			& 1110$\pm$110	& 1370$\pm$110	& 1460$\pm$390	& --		\\                
PKS 2356-61  		& \nodata			& \nodata				& 16$\pm$1.6$^\dagger$	& \nodata			& \nodata			& \nodata			& $>$10	\\
%--------------------------------------------------------------------------------------------------------------------------------------------------------------------------------------------------------------------------------
\enddata
\end{deluxetable*}

%%%%%%%%%%%%%%%%%%%%%%%%%%%%%%%%%%%%%%%%%%%%%%%%%%%%%%%%%%%%%%%%%%%%%%%%%%%%%%%%

\section{Spectral shapes and young radio sources fraction} \label{sec:GPS}

\begin{figure*}[tbp]
\begin{center}
\includegraphics[scale=0.60,angle=0]{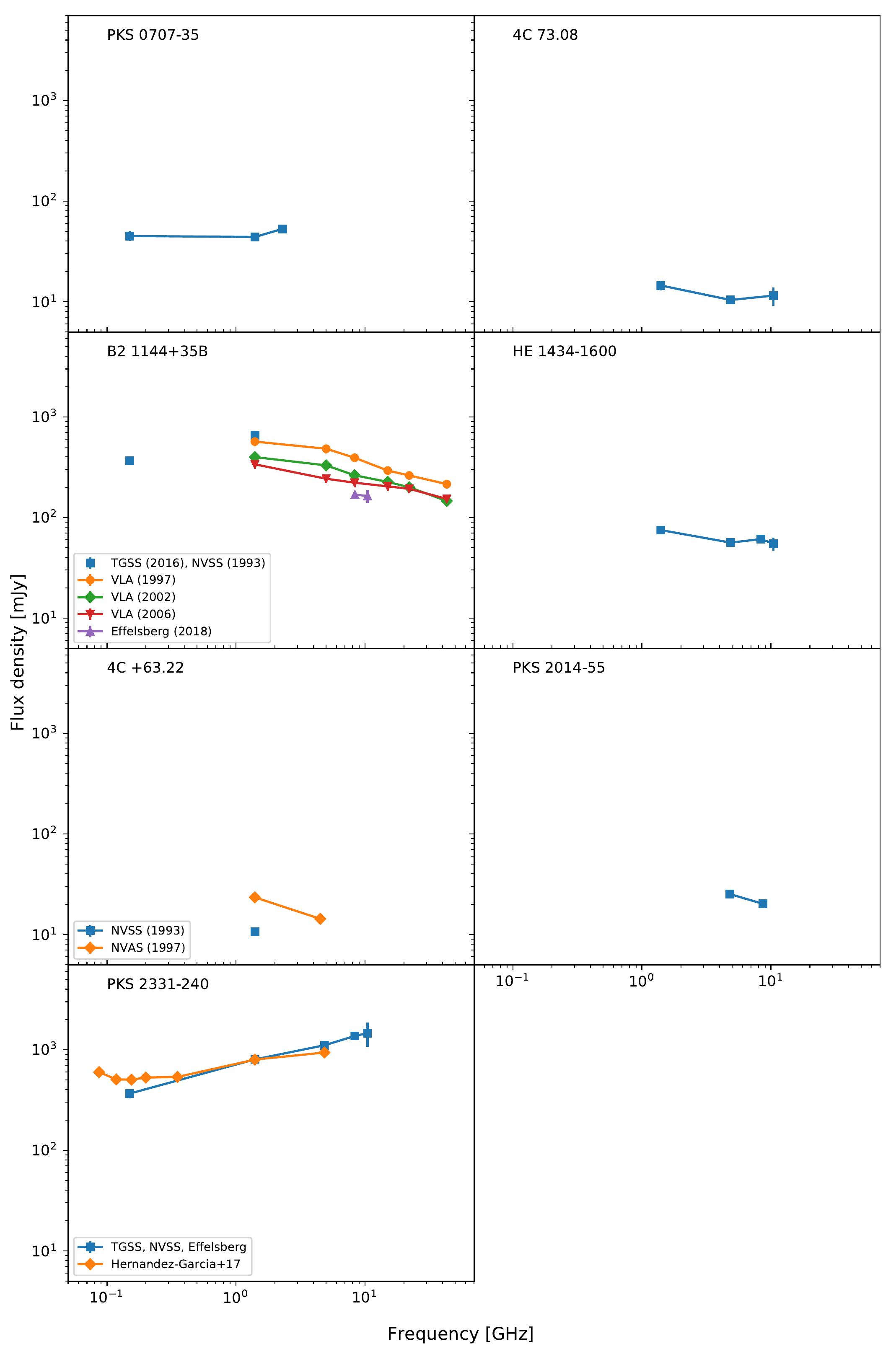}
\caption{Non-GPS radio cores spectra. Sources 4C +63.22 and PKS 2356-61 do not have enough spectral coverage to exclude the presence of a peak. Errors in logarithmic scale are smaller than symbol size for most of measurements. \label{fig:2}}
\end{center}
\end{figure*}

With the data described in the previous section, we were able to reconstruct the radio spectral shape of the fifteen sources of the sample, searching for the presence of a peak in the MHz-GHz frequency range. We have been able to gather at least 3 flux density measurements for 13 out of the 15 sources, allowing us to constrain the shape at least in the GHz range for the majority of them. In order to establish whether or not a GPS peak was present in the radio spectra of these sources, we performed a basic modeling of the data using a log-parabola as a simplified synchrotron emission component, analogously to previous works in the literature (\citealt{2012A&A...542A..13B, 2017MNRAS.467.4763T}). The procedure has the solely purpose of identifying the presence of a peak, and not to characterize the different free parameters of the synchrotron components expected from a core spectrum (and for which a better sampling would be required). In particular, we fitted the data with the Astropy {\tt LogParabola1D} function (\citealt{2018AJ....156..123A}), using the {\tt LevMarLSQFitter} routine that performs a Levenberg-Marquardt and least squares statistic. We excluded from the fitting procedure the points that were obviously part of a secondary, older, synchrotron component at low frequencies (see below for a discussion). We obtained a good fit for 6 objects (0318+684, IGR J14488-4008, Mrk 1498, 4C +34.47, IGR J17488-2338, 4C +74.26). For B3 0309+411B and PKS 2356-61, data above 1 GHz show an inverted spectrum, possibly indicating a self-synchrotron absorption, that could be fitted by a power-law with spectral index $\alpha$=0.6 and $\alpha$=0.5 respectively (using the {\tt PowerLaw1D} function), so we can assume a peak may be present at tens of GHz, beyond the spectral window sampled in this work. These could well be examples of HFP sources, thus even younger than the GPS ones. Peak frequency for the modeled GPS components are given in Tab. \ref{fluxes}. Four sources among the previous ones (0318+684, IGR J14488-4008, IGR J17488-2338, 4C +74.26) show also a convex spectrum, with a minimum at low frequencies with respect to the peak in the GHz range: \cite{2017ApJ...836..174C} presented a study of sources peaking at low frequencies ($<$1.4 GHz) making use of the GLEAM survey (\citealt{2017MNRAS.464.1146H}). They propose that convex-spectrum sources are undergoing a re-starting phase, responsible for the peak, while the lower frequency steep emission should be the remnant of a previous activity. %This further confirms the restarting scenario proposed above for our GPS/HFP sources, given that they lie at the center of GRG galaxies.
Plots of the successfully fitted spectra are presented in Fig. \ref{fig:1}, while the ones not showing a peak, or not having enough measurements for modeling (4C +63.22 and PKS 2014-55) are shown in Fig. \ref{fig:2}. 

Considering the 8 sources above, presenting hints of GPS or HFP peaks, and excluding the two sources for which the spectral coverage does not allow us to verify the presence of a peak (4C +63.22 and PKS 2014-55), we obtain a fraction of young radio sources of $61^{+30}_{-21}$\% (Poissonian errors are considered due to the small number of observed events, \citealt{1986ApJ...303..336G}). 
%%%%%%%%%%%%%%%%%%%%%%%%%%%%%%%%%%%%%%%%%%%%%%%%%%%%%%%
Given that these radio cores reside in old radio galaxies as implied by the large  dimensions of their lobes, this high fraction is surprising but difficult to interpret. It is in fact hard  to compare this fraction with those seen in other samples given the peculiar selection of our sources. Nevertheless we recall that the frequency of occurrence of the GPS and CSS sources in different flux density limited samples ranges from 8 to 10\% in the first case and  10-30\% in the second case (depending on the selection frequency, \citealt{1998PASP..110..493O} and references therein). More recently, \cite{2014MNRAS.438..796S} examining the radio population at 20GHz (a frequency dominated by the core emission)  suggests that the combined  fraction of GPS/CSS objects can be even higher, reaching values up to 40\%, if low luminosity sources are taken into account. On the other hand \cite{2017ApJ...836..174C}, studying the GLEAM radio source population, found that 4.5\% of the objects are peaked spectrum sources (GPS/HFP/CSS), thus indicating a large spread  in the observed fractions. Furthermore, one must underline that  the identification of these radio  sources is not straightforward, since on one hand it requires sufficient spectral data that a peak in the spectrum can be identified, while on the other hand it is typically limited to the analysis of unresolved,  i.e. compact,  objects. Thus, the  statistical values quoted above are probably lower limits.
 
To the best of our knowledge, systematic searches for CSS/GPS/HFP objects in the core of extended radio galaxies have never been performed, although a few examples of such objects were previously reported in the literature \citep{1998PASP..110..493O}; this is understandable, since FRI/FRII are thought to be the final stage of peaked spectrum radio sources, but the issue may be reconsidered in view of the evidence that recurrent activity in radio galaxies is indeed observed. Furthermore, such searches should be targeted to radio-selected samples of radio galaxies, whereas our sample is hard X-ray selected and thus likely to pick up objects with active and thus possibly young cores.

Considering our own sample of hard X-ray selected radio galaxies from \cite{2016MNRAS.461.3165B}, and limiting the search to objects with small dimensions - 36 with dimension below 200 kpc, assumed as maximum size for intermediate radio galaxies between CSS ($<$20 kpc) and the most extended ones ($>$200 kpc) - we could provide a first comparison sample, despite broad band radio data are scarse above 1 GHz, and  the AGN core is often radio weak or unresolved with respect to the large radio structure. Indeed, a detailed analysis of the entire sample would require an ad hoc observing campaign and is therefore beyond the reach of the present work. Nevertheless, searching the available literature, we were able to obtain spectral information for 12 of the 36 low size objects: 2 can be considered to be peaked-spectrum objects in the range 150 MHz - 10 GHz, thus similar to the one explored in this work (4C 50.55, \citealt{2007MNRAS.382..937M}; 3C 390.3, \citealt{1996A&A...308..376A}) while the other 10 have instead flat or steep spectrum cores (Cygnus A, \citealt{1996A&ARv...7....1C}; 3C 309.1, \citealt{1998PASP..110..493O}; Pictor A, \citealt{1997A&A...328...12P}; 3C 389, \citealt{2017ApJS..230....7P}; 3C 382 and 3C 452, \citealt{1973MNRAS.164..271R}; NGC 1275, 3C 120, 3C 184.1, and 3C 227, \citealt{2007ApJS..171...61H}). This suggests that young cores in our sample can be found in extended radio galaxies of all sizes, but are probably less common in small-sized objects (albeit their fraction has to be properly estimated), giving a first indication that the hard X-ray selection should not have a dominant effect on the abundance of such sources. Nevertheless, only a dedicated observing campaign could exclude it.

All together, and limiting the analysis to our GRG, we find that their cores are mostly young ($\sim$kyr) at radio frequencies while their structure is old and evolved ($\sim$Myr). From the analysis of their X-ray spectra \citep{2018MNRAS.481.4250U}, we also find that the nuclei of these GRG are presently active and powered by a hot corona coupled to an efficient accretion disc. Interestingly, a high energy cut-off has been measured in four X-ray spectra, and
all these four sources belong to the GPS class. The presence of a high-energy cut-off is a signature of the X-ray
emission originating via thermal Comptonization rather than synchrotron and/or inverse Compton in a jet, implying that the nuclear X-ray emission is not dominated by the jet but rather by the coronal emission of the active nucleus. Finally, a morphological study of the sample (\citealt{2019arXiv190201657B}, Bruni et al. in prep.) as well as the analysis of individual sources (PKS 2331-240, \citealt{2017A&A...603A.131H}; Mrk 1498, Hern\'andez-Garc\'ia et al. in prep.) also highlight objects with evident sign of restarting activity, further increasing the number of GRG from \cite{2016MNRAS.461.3165B} which experienced reactivation of their cores. This finding, together with the abundance of GRG in our sample, supports the scenario originally proposed by \cite{1996MNRAS.279..257S}, in which multiple episodes of activity would favor the growth of radio sources up to the extreme size of GRG.

%\cite{1996MNRAS.279..257S}
%\cite{2016MNRAS.461.3165B}

\subsection{Variability}

In general, the flux density of the sources in our sample is dominated by extended emission which is not variable, some amount of variability may be foreseen in the nuclear components in case the accretion is not constant, and geometrical effects (re-orientation of the radio axis) are favorable.
In fact, the symmetry and the size of the GRG implies that the large scale emission lies roughly in the plane of the sky. The nuclear region (often difficult to disentangle) instead may have been re-oriented. Furthermore, in our study, the resolution of the observations may force us to consider within the core region also portions of non-variable jet emission, which may partially wash out the core variability. With this caveat, 
we found significant variability for three of them (B2 1144+35B, 4C +63.22, 4C 34.47) in the core flux density from literature (B2 1144+35B, see previous section), or comparing archive and survey data. Indeed, 4C +63.22 show an increase of a factor of $\sim$2 at 1.4 GHz in the NVAS data with respect to NVSS (23.4$\pm$1.1 mJy versus 10.6$\pm$0.1 mJy) - NVAS data are from 1997, while NVSS from 1993. Analogously, for 4C 34.47 the two NVAS measurements at 15 GHz show a factor $\sim$3 decrease in 6 months, from November 1984 to May 1985 (767$\pm$38 mJy versus 226$\pm$11 mJy). This source was already studied by \cite{2010A&A...523A...9H}, that estimated a viewing angle of 57 degrees, and a core variability up to 25\% in ten years. Another two sources, B3 0309+411B and 4C 74.26, are known to have a variable core flux density from previous works: the first was presented by \cite{1984JApA....5..429S} and \cite{1989A&A...226L..13D}, who show an inverted spectrum up to 100 GHz with a variable core flux density, while the latter was investigated by \cite{1989MNRAS.236P..13R}, \cite{1992MNRAS.259P..13P}, and \cite{2004MNRAS.355..845K}, who found a variability of $\sim$50\% at 5 GHz over 14 years. 

Short-term flux density variability due to Doppler boosting is typical of sources with the jet base oriented within a few degrees from the observer's line of sight, although it can be found at larger angles ($\sim$40 degrees, \citealt{1989MNRAS.236P..13R}). A few percent variations can be also due to a synchrotron peak quickly moving towards lower frequencies, or to a newly ejected component. This could be the case for B3 0309+411, 4C 74.26, and 4C +34.47, that show an inverted or a GPS spectrum, while for 4C +63.22 we do not have enough information to draw any solid conclusion.

%%%%%%%%%%%%%%%%%%%%%%%%%%%%%%%%%%%%%%%%%%%%%%%%%%%%%%%%%%%%%%%%%%%%%%%%%%%%%%%%%%%%%%%%%%%%

\section{Conclusions} \label{sec:conclusions}

We carried out a radio observing campaign on a sample of 15 hard-X ray selected GRG, aiming at characterize the spectral shape of their cores from 150 MHz to 10 GHz, and estimate the fraction of young radio sources. The fraction of self-absorbed cores (either GPS or HFP-like, $61^{+30}_{-21}$\%) among the cores of these GRG suggests that a re-starting event is ongoing, given that these are located at the center of Mpc-scale - and thus evolved - radio structures. Together with the high fraction of GRG found by \cite{2016MNRAS.461.3165B} in their parent sample, these findings suggests that hard X-ray selected samples of radio galaxies present more easily Mpc-scale radio structures, and moreover their cores are often undergoing a restarting episode. 
This result underline the importance of systematic studies of the core spectra of evolved radio galaxies, in which a non-negligible fraction of reactivated nuclei could be present.

The capabilities of the next generation of radio (SKA, ngVLA) and X-ray (ATHENA) telescopes together with LOFAR, will make possible to confirm whether the fraction of restarted nuclei in radio galaxies has a dependence on hard X-ray luminosity or whether it is intrinsically high, through detailed studies of the link between radio and X-ray emission in these unique structures.

%%%%%%%%%%%%%%%%%%%%%%%%%%%%%% ACKNOWLEDGEMENTS %%%%%%%%%%%%%%%%%%%%%%%%%%%%%%%%%
\vspace{1cm}
GB acknowledges financial support under the INTEGRAL ASI-INAF agreement 2013-025-R1.
EC acknowledges  the  National  Institute  of  Astrophysics (INAF) and 
the University of Rome - Tor Vergata for the PhD scholarship in the 
XXXIII PhD cycle. 
LHG acknowledges support from FONDECYT through grant 3170527. 
This publication has received funding from the European Union's Horizon 2020 research and innovation programme under grant agreement No. 730562 [RadioNet]. 
We acknowledge support from a grant PRIN-INAF SKA-CTA 2016.
This work is partly based on observations with the 100-m telescope of the MPIfR in Effelsberg. We thank the staff of the GMRT that made these observations possible. GMRT is run by the National Centre for Radio Astrophysics of the Tata Institute of Fundamental Research.

%%%%%%%%%%%%%%%%%%%%%%%%%%%%%%%% BIBLIOGRAPHY %%%%%%%%%%%%%%%%%%%%%%%%%%%%%%%%%%%

\bibliographystyle{aasjournal}
\bibliography{references} % if your bibtex file is called example.bib

%%%%%%%%%%%%%%%%%%%%%%%%%%%%%%%%%% FIGURES %%%%%%%%%%%%%%%%%%%%%%%%%%%%%%%%%%%

%\newpage

%%%%%%%%%%%%%%%%%%%%%%%%%%%%%%%%%%%%%%%%%%%%%%%%%%%%%%%%%%%%%%%%%%%%%%%%%%%%%%%%%%%%%%%%%%%%%%%%



\end{document}